\documentclass[a4paper,11pt]{article}
\pdfoutput=1 

\usepackage{jinstpub} 

\usepackage{lineno}
\usepackage{subfig}

\title{A high frequency radiation hardened DC/DC-converter with low volume air core inductor}

\author[a,1]{J. Kampkötter,\note{Corresponding author.}}
\author[a]{M. Karagounis,}
\author[b]{R. Kokozinski}

\affiliation[a]{University of Applied Sciences and Arts Dortmund,\\Sonnenstr. 96-100, Germany}
\affiliation[b]{University of Duisburg,\\Germany}

\emailAdd{jeremias.kampkoetter@fh-dortmund.de}

\abstract{The hfrh-buck (high frequency radiation hardened-buck) is a radiation hardened DC/DC-converter operating at a high switching frequency of 100MHz with a small air core inductor of 22nH. To ensure a high radiation dose, the circuit is designed with core transistors of a 65nm TSMC technology. By stacking the transistors of the power stage, the converter can be supplied with a voltage of up to 4.8V. Stable operation can be achieved at an output voltage of 1.2V with a maximum load current of 1A. The prototype demonstrates the ability to power parallel connected hybrid-pixel modules in the innermost layers.}


\proceeding{Topical Workshop on Electronics for Particle Physics 2023\\
  02-06 October, 2023\\
  Sardinia, Italy}

\begin{document}
\maketitle
\flushbottom

\section{Introduction}
The readout electronics and voltage regulators of the pixel detector are consistently improving to ensure a reliable operation over the lifetime of the experiments and to meet the demands of harsh conditions. To minimize the weight of detector systems, the voltage regulators are designed in the same technology as the detector readout chip, enabling on-chip integration. The voltage regulators must reliably operate under high radiation doses, limited volume, long operating times, and intense magnetic fields. Due to these challenging demands, the development of innovative power supply concepts for high energy physics applications is of great interest. The proposed DC/DC-converter has the ability to power future hybrid pixel modules in parallel even in harsh environments. 

\section{System Overview}
\begin{figure}[htbp]
\centering 
\includegraphics[width=.7\textwidth]{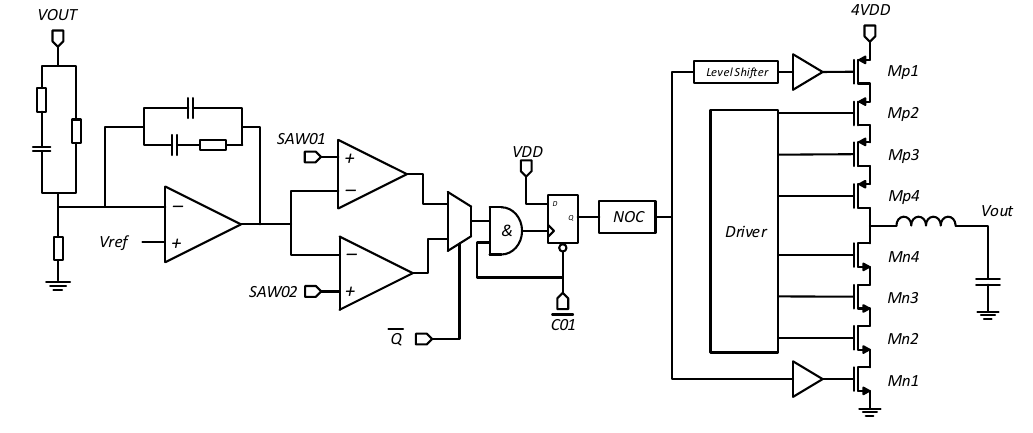}
\caption{\label{fig:sysOV}DC/DC-converter with stacked switching stage and a PWM voltage mode regulator}
\end{figure}

This section provides insights into the design of the DC/DC-converter, which is capable of withstanding high radiation doses of up to 1 Grad, enabling operation in the innermost layers of a detector system. The entire circuit is built with low-voltage, thin-gate oxide core transistors in TSMC 65nm CMOS technology, as this has shown high radiation tolerance. However, to accommodate high supply voltages that require thick gate oxide transistors, the switching stage is designed using stacked core transistors to achieve both objectives. Due to the high magnetic fields of up to several Tesla in the experiments, ferromagnetic cores cannot be used because they are saturated and lose their positive properties. Similarly, bulky air coils are unsuitable due to space constraints. Thus, a switching frequency of 100MHz was selected to decrease the necessary inductance to only 22nH. The DC/DC-converter is based on a conventional buck topology with a PWM voltage mode control for output regulation, as shown in figure \ref{fig:sysOV}.

\subsection{Device Stacking}
The switching stage is built with four PMOS and four NMOS devices to accommodate a regulator input voltage that is four times higher than the rated supply voltage of a single transistor. The cascode switching circuit is a crucial component because it must carry high currents reliably with fast switching times. Additionally, it is important to ensure that the stacked transistors operate within their voltage limits to prevent destruction or accelerated aging. The core transistors typically have a maximum voltage rating of 10\% above their nominal voltage, resulting in a small voltage margin. To alternate between the two switching states, transistors Mp1 and Mn1 are driven by two control signals. In addition, a bias network \cite{a} is selected to apply the necessary gate voltages to the cascode transistors by monitoring the drain voltage of the power devices. In \cite{b}, the driver network with power stage was analyzed for use in a high frequency DC/DC-converter and an improved concept was presented to avoid hot carrier effects. Tapered buffers drive the power devices with very fast switching times in the range of 100ps. To prevent crosstalk currents, a non-overlapping clock generator was implemented, to ensure that the pull-up and pull-down stages do not conduct simultaneously. 

\subsection{Analog building blocks for fast switching operation}

Fast switching times, small duty cycles, and high load currents are a significant challenge in designing reliable analog blocks. To regulate the output voltage, a PWM voltage mode control with type 3 compensation was selected. The feedback network must meet high demands to ensure reliable operation at high switching frequencies of 100MHz with simultaneous low duty cycles of 0.25. Several techniques have been implemented to guarantee safe operation and are exhibited in the following sections. 

\subsubsection{Sawtooth Generator}
The sawtooth generator is designed to generate two triangular waveforms in order to ensure fast switching times with small duty cycles \cite{c}. An active current source charges and discharges two capacitors alternatively, as illustrated in figure \ref{fig:sawosc}. During this operation, the bias transistor must remain in saturation, thus defining an upper and lower threshold voltage as $V_{HI} = V_{DD} - V_{DSAT}$ and $V_{LO} = V_{DSAT}$. The waveforms in the form of a triangle are observed by comparators and compared with the specified threshold voltages. When the capacitor to be charged reaches the upper threshold, the clock signal of the D-flipflop is triggered, causing the transmission gates to switch on or off. As a result, the second capacitor is charged and the first capacitor is discharged. The discharge process ceases when the lower threshold is reached. The primary benefit of this method is that the circuit shifts to the second capacitor that is depleted when the first capacitor is charged. This prevents the discharging process from constraining the duty cycle, as is the case when only one capacitor is used. Additionally, fast falling edges can be evaded, which reduces the impact of crosstalk at the comparator input.

\subsubsection{Level Shifter}
\begin{figure}[h]
  \centering
  \subfloat[]{
    \includegraphics[width=0.45\linewidth]{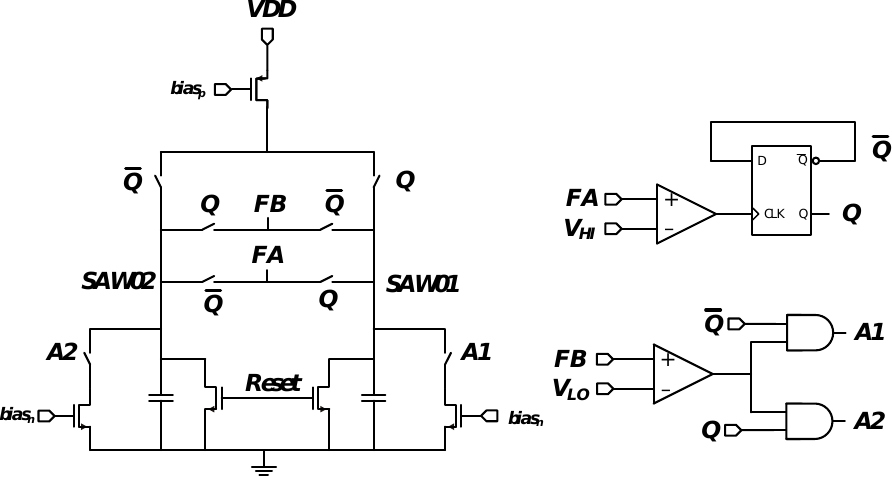}
    \label{fig:sawosc}
    }
  \qquad
  \subfloat[]{
    \includegraphics[width=0.45\linewidth]{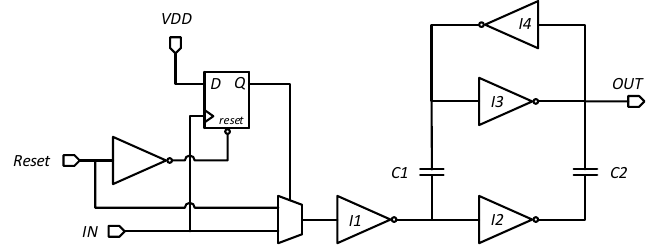}
    \label{fig:LevelShifter}
  }
  \caption{a) Overview of the sawtooth oscillator. b) Capacitive floating level shifter with reset}
\end{figure}

A level shifter is required to shift the control signal to the high-side domain in order to drive the PMOS power switch Mp1. A capacitive floating level shifter \cite{d} was chosen as shown in figure \ref{fig:LevelShifter}. The level shifter comprises  two inverters configured in positive feedback, forming a latch. Coupling capacitors enable the transition from the low-side to the high-side domain and facilitate the control signal transmission between the two voltage levels. The capacitive floating level shifter usually has a straightforward design that requires only a few components. With small capacitors in the femtofarad-range, the required chip area is also small. The design approach outlined in the work by \cite{e} was selected to ensure a stable latch flip during low side signal domain edges. A reset mechanism is implemented to guarantee the proper starting state of the latch.

\subsubsection{PWM Generation Concept}

\begin{figure}[!tb]
  \centering
  \subfloat[]{
    \includegraphics[width=0.56\linewidth]{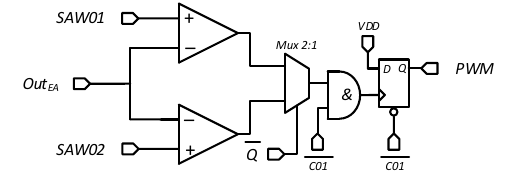}
    \label{fig:feedback}
    }
  \quad
  \subfloat[]{
    \includegraphics[width=0.38\linewidth]{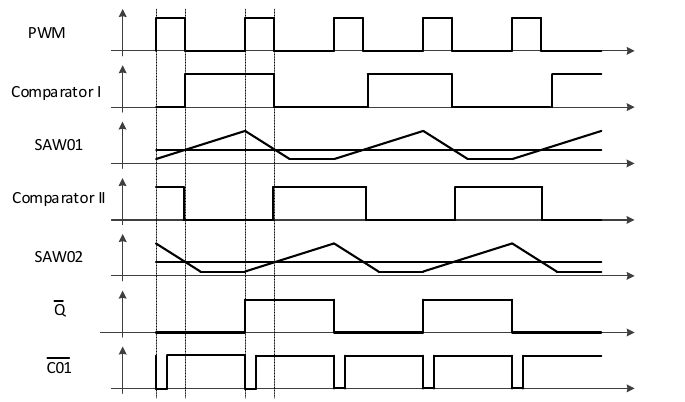}
    \label{fig:waveform}
  }
  \caption{a) PWM generation circuit. b) Waveforms for generation of PWM signal}
\end{figure}

The PWM signal is generated using a concept based on two comparators. The sawtooth voltages $SAW01$ and $SAW02$ are phase shifted by one switching period and connected to their respective comparator input. When one of the sawtooth voltages rises, the multiplexer selects the corresponding comparator output to define the PWM control signal. At the same time, the second sawtooth voltage slowly decreases as the corresponding capacitor in the oscillator discharges. The $\overline{Q}$ signal displayed in figure \ref{fig:sawosc} alternates with each switching period and is responsible for selecting which of the multiplexer inputs are forwarded to the switching stage. The timing diagram is shown in figure \ref{fig:waveform}. As soon as $SAW01$ or $SAW02$ reaches the upper threshold voltage $V_{HI}$, the D-flipflop is reset by the control signal $\overline{C01}$, as illustrated in figure \ref{fig:feedback}, and the PMOS devices become conductive. When one of the waveform voltages exceeds the output of the error amplifier, the PWM signal goes high, causing all NMOS devices to become active and pull the switching node Vx to ground. This methodology permits the capacitors of the sawtooth generator to discharge with a slow falling edge, which helps reduce crosstalk effects of the comparator. At the start of a new switching period, the D-flipflop guarantees that the PMOS stage is closed. An additional AND gate effectively shuts down the NMOS stage once the error amplifier output rises beyond the upper sawtooth threshold. This prevents the circuit from a permanently closed pull-up stage due to an absent edge. The proposed concept ensures a robust operation with reliable switching operation.

\section{Experimental results}

\begin{figure}[h]
  \centering
  \subfloat[]{
    \includegraphics[width=0.4\linewidth]{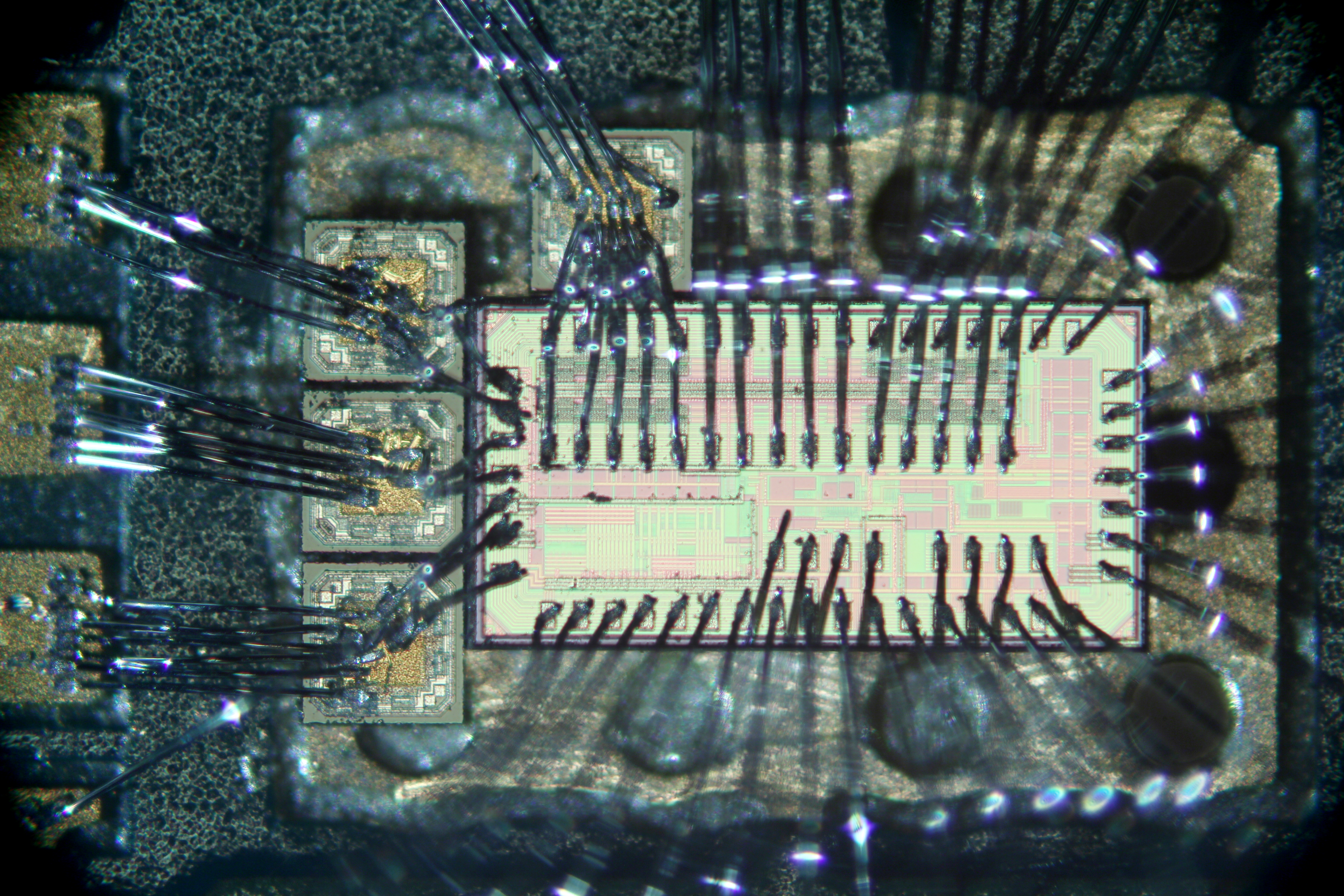}
    \label{fig:chip}
    }
  \qquad
  \subfloat[]{
    \includegraphics[width=0.48\linewidth]{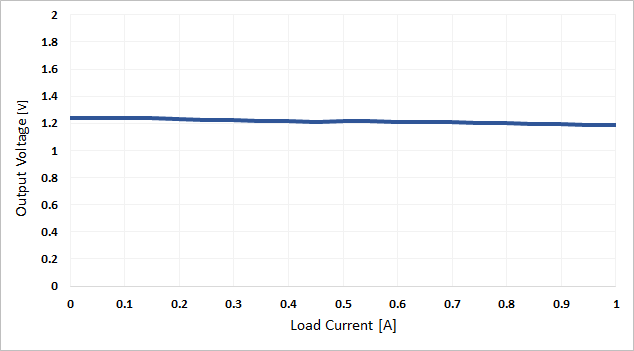}
    \label{fig:loadreg}
  }
  \caption{a) Chip micrograph with silicon decoupling cap. b) Measurement of load regulation}
\end{figure}

The proposed DC/DC-converter has been implemented in 65nm TSMC CMOS technology as shown in the chip micrograph in figure \ref{fig:chip}. An air core inductor of 22nH and the output capacitor are placed off-chip. On-chip decoupling capacitors, ranging from 100 to 200pF, have been implemented to mitigate oscillations at the four critical supply voltages because of the fast switching edges and low device voltages. Connecting ten bond wires in parallel reduces the total inductance of the bond wire for the 4VDD supply voltage. To further decrease high input ripple, bondable silicon capacitors have been strategically placed adjacent to the bare die. This not only shortens the bondwire length but also minimizes the influence of the PCB trace inductance, since the decoupling capacitor is in close proximity to the die. Also, the ESL of the output capacitor has a dominant influence on the operation of the regulator due to the fast switching edges. Therefore, several capacitors with different capacitance values are arranged in parallel in close proximity to the chip in order to lower the amount of output voltage ripple. 

\begin{table}[htbp]
\centering
\begin{tabular}{|l|c|}
\hline
Supply Voltage / Output Voltage & 4.8V / 1.2V \\
\hline
Switching Frequency & 100MHz \\
\hline
Max. load current & 1A \\
\hline
Inductor and Capacitor & 22nH / 100nF \\
\hline
Peak Efficiency @ 400mA & 70\% \\
\hline
\end{tabular}
\caption{\label{tab:Over} Testchip overview.}
\end{table}

\begin{figure}[h]
  \centering
  \subfloat[]{
    \includegraphics[width=0.3\linewidth]{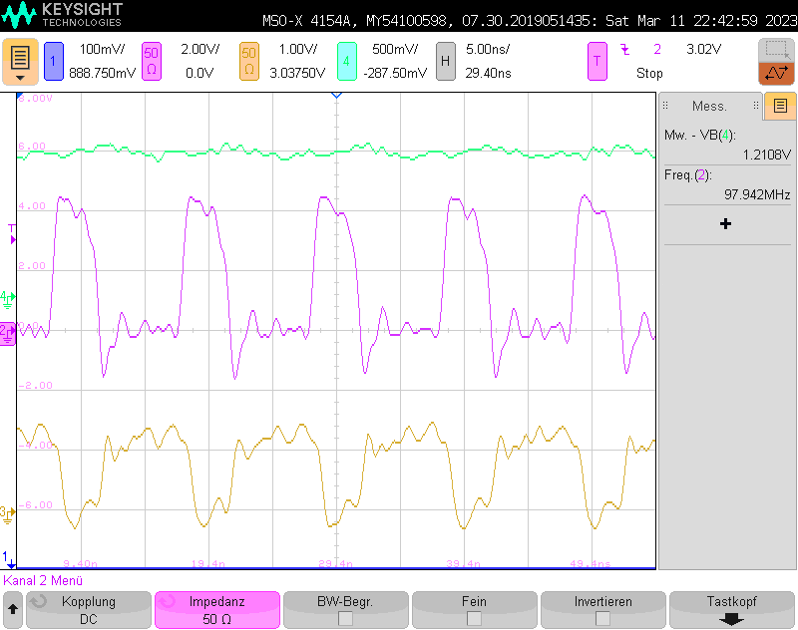}
    \label{fig:tranMES}
    }
  \quad
  \subfloat[]{
    \includegraphics[width=0.3\linewidth]{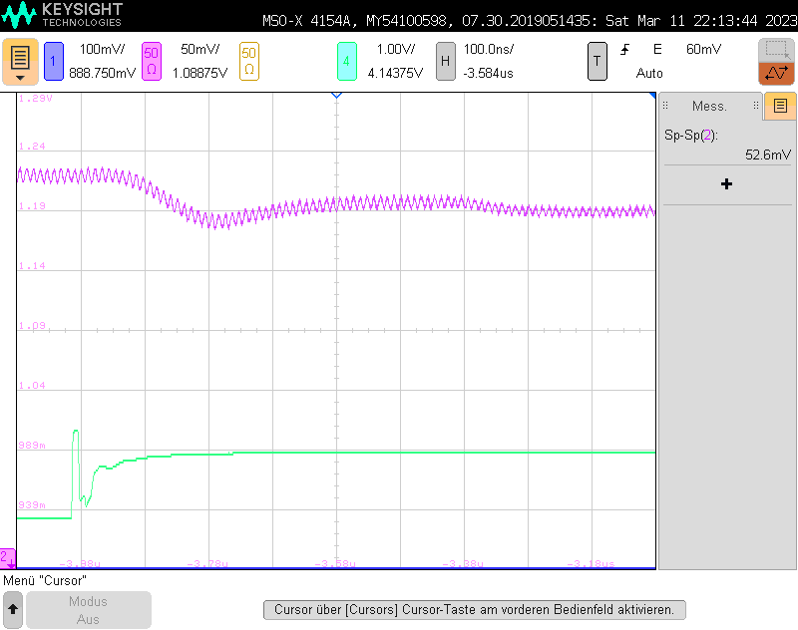}
    \label{fig:STEPC01}
  }
   \quad
  \subfloat[]{
    \includegraphics[width=0.3\linewidth]{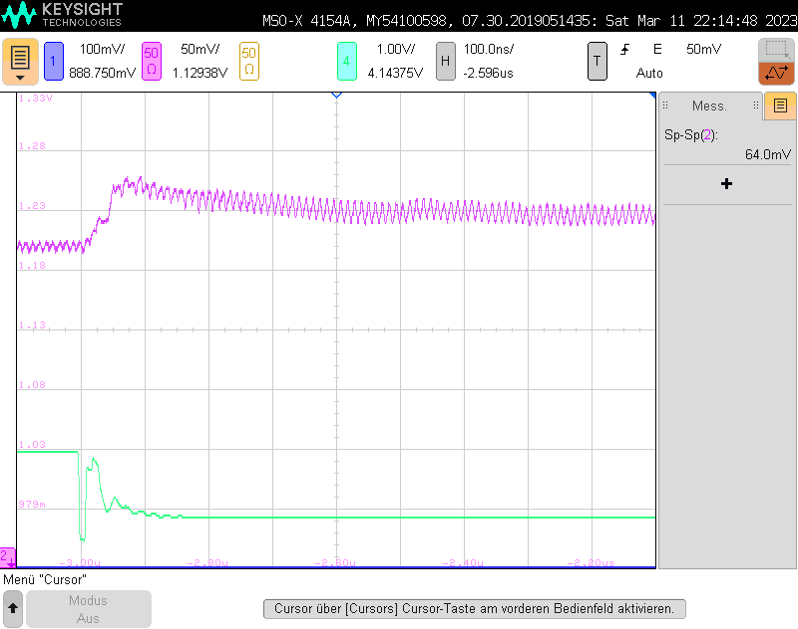}
    \label{fig:STEPC02}
  }
  \caption{a) Transient Measurement. b) Step load change from 0A to 0.35A. c) Step load change from 0.35A to 0A}
\end{figure}

Table \ref{tab:Over} presents the prototype specifications. A waveform in steady state, as shown in figure \ref{fig:tranMES}, exhibits stable operation under a load current of 500mA with a supply voltage of 4.8V. The output voltage marked in green is regulated to the desired voltage of 1.2V. The rectangular switch node voltage Vx shown in purple switches at 100MHz and the control signal of NMOS device Mn8 is shown in yellow. A load regulation shown in figure \ref{fig:loadreg} was carried, to examine the output voltage across the entire load range. Sudden load changes from 0A to 0.35A and back from 0.35A to 0A have also been investigated and show a fast settling time of 700ns and 500ns respectively as can be seen in figure \ref{fig:STEPC01} and \ref{fig:STEPC02}. A peak efficiency of 70\% is attained at a load current of 400mA.

\section{Conclusion}
A high frequency DC/DC converter with a stacked switching stage is introduced, for powering future high-energy physics experiments. The converter is built using low-voltage core devices, which can endure high doses of irradiation, because the core transistors are radiation tolerant due to their thin gate oxide. By stacking the devices of the switching stage, the DC/DC-converter can be supplied with a four times higher supply voltage. Due to the restricted space and strong magnetic fields in the experiments, the converter operates at a switching frequency of 100MHz and requires a small 22nH air core inductor. A number of methods were proposed to ensure dependable operation with a maximum load current of 1A. The results demonstrate that the prototype can operate in a harsh environment and meet the power efficiency requirements with low mass penalty.

\appendix

\end{document}